\documentclass[useAMS,usenatbib]{mn2e}

 \usepackage{times}
 \usepackage{graphicx}

\newbox\grsign \setbox\grsign=\hbox{$>$} \newdimen\grdimen
\grdimen=\ht\grsign
\newbox\simlessbox \newbox\simgreatbox \newbox\simpropbox
\setbox\simgreatbox=\hbox{\raise.5ex\hbox{$>$}\llap
     {\lower.5ex\hbox{$\sim$}}}\ht1=\grdimen\dp1=0pt
\setbox\simlessbox=\hbox{\raise.5ex\hbox{$<$}\llap
     {\lower.5ex\hbox{$\sim$}}}\ht2=\grdimen\dp2=0pt
\setbox\simpropbox=\hbox{\raise.5ex\hbox{$\propto$}\llap
     {\lower.5ex\hbox{$\sim$}}}\ht2=\grdimen\dp2=0pt
\def\simgreat{\mathrel{\copy\simgreatbox}}
\def\simless{\mathrel{\copy\simlessbox}}

\title[Praesepe white dwarfs]{High resolution optical spectroscopy of Praesepe white dwarfs \thanks{Based on observations made with ESO telescopes at the La Silla Paranal Observatory under programme ID 076.D-0751}}
\author[S. L. Casewell et al.]{S. L. Casewell$^{1}$\thanks{E-mail:
slc25@star.le.ac.uk}, P. D. Dobbie$^{2}$, R. Napiwotzki$^{3}$, M. R. Burleigh$^{1}$,
M. A. Barstow$^{1}$, \newauthor R. F. Jameson$^{1}$
\\
$^{1}$Department of Physics and Astronomy, University of Leicester, University Road, Leicester LE1 7RH, UK \\
$^{2}$Anglo-Australian Observatory, P.O. Box 296, Epping 1710, Australia \\
$^{3}$ Science \& Technology Research Institute, University of Hertfordshire,
College Lane, Hatfield, AL10 9AB \\
}

\begin{document}

\date{Accepted \today{}. Received \today{}; in original form \today{}}

\pagerange{\pageref{firstpage}--\pageref{lastpage}} \pubyear{2008}

\maketitle

\label{firstpage}

\begin{abstract}

We present the results of a high resolution optical spectroscopic study of nine white dwarf candidate members of 
Praesepe undertaken with the VLT and UVES. 
We find, contrary to a number of previous studies, that WD0836+201 (LB390, EG59) 
and WD0837+199 (LB393, EG61) are magnetic and non-magnetic white dwarfs respectively. Subsequently, we determine the 
radial velocities for the eight non-magnetic degenerates and  provide compelling evidence
that WD0837+185 is a radial velocity variable and possibly a double-degenerate system. We also find that our result for WD0837+218,
 in conjunction with its projected spatial location and position in initial mass-final mass space, argues it is more likely to be a field star than
a cluster member.
After eliminating these two white dwarfs, and WD0836+199 which has no clean SDSS photometry, we use the remaining 5 stars to substantiate modern theoretical mass-radius relations for white dwarfs.  In light of our new results we re-examine the white dwarf members of Praesepe and use them to further constrain the 
initial mass-final mass relation. We find a a near monotonic IFMR, which can still be adequately represented by simple linear function with only one outlier which may have formed from a blue straggler star. 

\end{abstract}


\section{Introduction}


The initial mass-final mass relation (IFMR) is a theoretically predicted positive correlation between the main sequence mass of a star
with M$\simless$10 M$_{\odot}$ and the mass of the white dwarf remnant left behind after it has expired (e.g. \citealt*{iben83}). 
Understanding the form of this relation is important since it provides a handle on the total amount of gas enriched with He, N and other
metals that low or intermediate mass stars, which account for 95  per cent of all stars, return to the interstellar medium at the end of their lifecycles.
Moreover, the form of the upper end of the IFMR is relevant to studies of Type II supernovae as it can provide a constraint on the 
minimum mass of star that will experience this fate.

The form of the IFMR is extremely difficult to predict from theory alone due to the many complex processes occurring during the final phases
of stellar evolution (e.g. third dredge-up, thermal pulses, mass loss; \citealt*{iben83}). This means that robust empirical data are essential for constraining 
its form. However, these are by no means simple to obtain, a significant difficulty being the determination of the main sequence mass of a
star that has long since ceased to exist. This difficulty can be alleviated by using white dwarf members of open star clusters 
\citep{bergeron95, claver01, dobbie04, williams04, kalirai05, dobbie06, dobbie06b, kalirai07, williams07} to define the IFMR (e.g.\citealt{weidemann77,weidemann00, ferrario05, dobbie06, williams07, kalirai08}). Here, since the age of the population can be determined from the location of the main sequence
turn-off (\citealt{sandage56}; e.g. \citealt{king05} in Ursa Major) the lifetime, and ultimately the mass, of the progenitor star of any degenerate member can be estimated by calculating 
the difference between the cooling time of the white dwarf and the cluster age.

Nonetheless, until relatively recently, rather few white dwarf members of open star clusters had been identified. \citet{dobbie06} suggest $\sim$30 associated with 14 clusters, while the IFMR derived by \citet{ferrario05} used 40 DA white dwarfs from 7 open star clusters, including NGC2099. The uncertainties in membership status, cluster ages and the relatively large distances of these scarce, faint objects resulted in large scatter in 
the IFMR \citep{claver01, ferrario05}. To begin to address this issue and gauge the level of intrinsic scatter in the IFMR, we recently investigated the 
white dwarf members of the moderately rich nearby Praesepe open cluster. At a distance of 177$^{+10.3}_{-9.2}$ pc (as determined from Hipparcos 
measurements, \citealt{mermilliod97}), it is one of the closest star clusters. It is slightly metal rich with respect to the Sun ([Fe/H] = +0.11, \citealt{an07}).  Indeed, as both the metallicity and the kinematics of Praesepe are similar to those of the Hyades, the former is often touted as a 
member of the Hyades moving group and therefore is assumed have an age comparable to the latter, $\tau$=625$\pm$50 Myr (e.g. \citealt{claver01}). We note that this age for the Hyades was derived by comparing model isochrones generated from slightly metal enhanced (Z=0.024) stellar models which included moderate convective overshooting
to the colours and magnitudes of a sample of cluster members selected using Hipparcos astrometric data (\citealt{perryman98}). 

In \citet{dobbie04,dobbie06} we increased the number of probable white dwarf members of Praesepe from five to eleven. All these candidate members share 
the proper motion of the cluster (e.g. \citealt{claver01,dobbie04}). We used estimates from the literature and our own measurements 
of the effective temperature and the surface gravity of these white dwarfs to determine their masses and the masses of their progenitor stars so that 
we could study their locations in initial mass-final mass space. Combining these with data from a number of other open clusters with comparatively
well constrained ages and the Sirius binary system, we concluded that most stars follow rather closely a monotonically increasing IFMR, which can be
approximately described by a simple linear function over the initial mass range  2.7 M$_{\odot}$ to 6 M$_{\odot}$. Nevertheless, there existed a small number of outliers, 
all attributable to the Praesepe cluster.

Here we present the results of a high resolution optical spectroscopic study of nine out of eleven candidate Praesepe degenerates. In the next section we 
describe the data acquisition and analysis, including the measurement of line core velocity shifts and the redetermination of effective temperatures and 
surface gravities. We then utilise the five most appropriate Praesepe degenerates to re-examine the theoretical white dwarf mass-radius relation. Subsequently
we determine the radial velocities for all eight non-magnetic white dwarfs and use these to confirm their membership status. Finally, we examine our new 
results in the context of the IFMR.

\begin{figure*}
\begin{center}
\scalebox{0.50}{\includegraphics[angle=270]{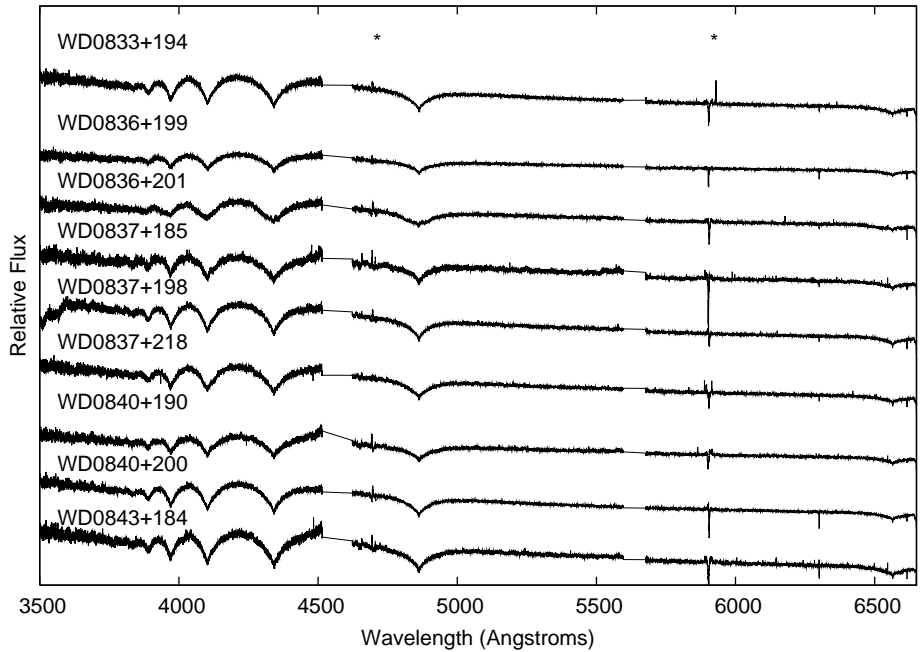}}\\
\caption{\label{wds} The full median filtered, smoothed, stacked UVES spectra for the nine white dwarfs included in 
this study. Note the Zeeman shifted components of the Balmer lines in the spectrum of WD0836+201 (see Figure \ref{wd0836+201} for more detail). The asterisks mark artifacts caused by dead pixels in the array. There are also two breaks in the data of width $\sim$80\AA\ at 4580 and 5640\AA\ caused by the edges of the CCDs.}
\end{center}
\end{figure*}

\begin{table*}
\caption{\label{teff}ID, effective temperature, log~$g$, mass and cooling time for the Praesepe white dwarfs with UVES spectroscopy. 
The previous effective temperature and surface gravity estimates are from \citet{dobbie06} and \citet{claver01}, where the values shown 
are as listed in this literature. The FITSB2 data are shown with the formal fitting errors. An uncertainty of 2.3 per cent and 0.07 dex are more realistic for $T_{\rm eff}$ and log~$g$.
Additionally, for Praesepe members, we list our estimate of the progenitor mass. The magnetic white dwarf WD0836+201 has been omitted as our models do not include a treatment for magnetism. Additionally, no initial mass has been given for WD0837+218 as it is believed more likely to be a field star than a cluster member.}
\begin{center}
\begin{tabular}{lllllcccc}
\hline
ID &$T_{\rm eff}$ &log~$g$ & FITSB2 $T_{\rm eff}$  & FITSB2 log~$g$ & M & R & $\tau$$_{\rm cool}$ & M$_{\rm init}$\\ 
&K&&K& & M$_{\odot}$ &$\times$10$^{-2}$ R$_{\odot}$ & Myrs & M$_{\odot}$\\
\hline
WD0833+194&     14999 $^{+208}_{-258}$& 8.18$^{+0.04}_{-0.03}$  & 14852$\pm$41 &8.182$\pm$0.005    &  0.721$\pm$0.043 & 1.143$\pm$0.103 &  270$^{+32}_{-28}$  & 3.32$^{+0.33}_{-0.22}$  \\
WD0836+199&     14060 $\pm$630        & 8.34$\pm$0.06           & 14571$\pm$60   &8.233$\pm$0.010  &  0.752$\pm$0.044 & 1.102$\pm$0.099 &  308$^{+37}_{-32}$  &  3.46$^{+0.43}_{-0.27}$  \\
WD0837+185&     14748$^{+396}_{-404}$ & 8.24$^{+0.06}_{-0.05}$  & 15076$\pm$60 &8.306$\pm$0.010    &  0.804$\pm$0.044 & 1.039$\pm$0.094 &  319$^{+39}_{-35}$  & 3.50$^{+0.48}_{-0.29}$ \\
WD0837+199&     17098$\pm$350         & 8.32$\pm$0.05           & 17240$\pm$38   &8.195$\pm$0.006  &  0.737$\pm$0.043 & 1.130$\pm$0.102 &  179$^{+23}_{-20}$  & 3.07$^{+0.20}_{-0.15}$ \\
WD0837+218&     16833$^{+236}_{-272}$ & 8.39$^{+0.04}_{-0.02}$  & 16875$\pm$44   &8.473$\pm$0.008  &  0.909$\pm$0.044 & 0.919$\pm$0.083 &  305$^{+40}_{-35}$  &  - \\ 
WD0840+190&     14765$^{+264}_{-277}$ & 8.21$^{+0.03}_{-0.03}$  & 14935$\pm$68   &8.381$\pm$0.010  &  0.849$\pm$0.045 & 0.985$\pm$0.089 &  368$^{+46}_{-40}$  & 3.73$^{+0.71}_{-0.39}$\\
WD0840+200&     14178 $\pm$350        & 8.23$\pm$0.05           & 14983$\pm$42   &8.176$\pm$0.005  &  0.721$\pm$0.043 & 1.143$\pm$0.103 &  263$^{+31}_{-28}$  & 3.30$^{+0.32}_{-0.21}$\\
WD0843+184&     14498$^{+199}_{-206}$ & 8.22$^{+0.04}_{-0.04}$  & 15018$\pm$50   &8.340$\pm$0.008  &  0.823$\pm$0.045 & 1.016$\pm$0.092 &  339$^{+42}_{-37}$  & 3.59$^{+0.55}_{-0.33}$\\

\hline
\end{tabular}
\end{center}
\end{table*}

%
\section{Acquisition and reduction of data}
\label{data}
High resolution spectroscopy of nine of the eleven DA white dwarf candidate members of Praesepe was obtained using the Ultraviolet and Visual 
Echelle Spectrograph (UVES) on the UT2 facility of the European Southern Observatory's (ESO) Very Large Telescope at Cerro Paranal, Chile. The 
data were acquired in service mode within the period 2005/11/16 and 2006/01/19 and using 45 minute exposure times. As only relaxed constraints were specified for the sky conditions 
these data were generally obtained through thin cirrus and in mediocre seeing ($\simless$1.4''). 

The observations utilised the UVES parallel mode, where cross dispersers CD2 and CD3, in conjunction with dichroic1 (390+580) and the HER5 and 
SHP700 filters provided simultaneous wavelength coverage over 370-500nm and 420-680nm in the blue and red arms respectively.  The 1.5'' slit 
resulted in a nominal spectral resolution of $\lambda$/$\Delta$$\lambda$$\sim$20000.

The data were reduced using the UVES pipeline which is based on the MIDAS software supplied by ESO \citep{ballester95}. The pipeline was run in OPTIMAL 
mode and an additional custom routine written by R. Napiwotzki applied to reduce the magnitude of the ripples which are sometimes apparent in these spectra. 
Calibration frames were taken every night science data was acquired. These include format-check frames, bias frames, flat fields, an order definition 
template, and a ThAr arc lamp spectrum. In addition to these calibrations, a spectrally featureless DC white dwarf (WD0000-345) was observed with the 
same instrument set-up on 2006/07/03. The spectrum of this star was used to remove remaining UVES instrumental signature from the data.  
The completed spectra, which each have a S/N$\sim$25 per resolution element, can be seen in figure \ref{wds}. These spectra have been median filtered, smoothed and co-added for the purpose of this figure.

%
\subsection{Data analysis}
\label{dataanal}
Our first step in analysing these echelle data was to remove the signature of telluric water vapour from around H$\alpha$ using an absorption
template constructed from numerous UVES spectra of white dwarfs obtained at Paranal Observatory as part of the SpY programme \citep{napiwotzki01}.
Next, the data were compared to the predictions of white dwarf model atmospheres using the spectral fitting programme FITSB2 (v2.04; \citealt{napiwotzki04}). 

The grid of pure-H model spectra was calculated using the plane-parallel, hydrostatic, non-local thermodynamic equilibrium (non-LTE) atmosphere code 
TLUSTY, v200 \citep{hubeny88,hubeny95} and the spectral synthesis code SYNSPEC v48 \citep{hubeny01}. TLUSTY assumes plane-parallel geometry and 
hydrostatic equilibrium. The models include a treatment for convective energy transport according to the ML2 prescription of \citet{bergeron92},
adopting a mixing length parameter, $\alpha$=0.6. These calculations utilised a model H-atom which incorporates explicitly the eight lowest 
energy levels and represents levels n=9 to 80 by a single superlevel. The dissolution of the high lying levels was treated by means of the occupation 
probability formalism of \citet{hummer88} generalised to the non-LTE atmosphere situation by \citet{hubeny94}. All calculations include the 
bound-free and free-free opacities of the H$^{-}$ ion and incorporate a full treatment for the blanketing effects of HI lines and the 
Lyman $-\alpha$, $-\beta$ and $-\gamma$ satellite opacities as computed by N. Allard \citep{allard04}. During the calculation of the model structure
the lines of the Lyman and Balmer series were treated by means of an approximate Stark profile but in the spectral synthesis step detailed profiles 
for the Balmer lines were calculated from the Stark broadening tables of \citet{lemke97}. The grid of model spectra covered the $T_{\rm eff}$ range of 
13000-20000 K in steps of 1000 K and log~$g$ between 7.5 and 8.5 in steps of 0.1 dex.

\subsection{Determination of effective temperature and gravity}
\label{dataanal}

Each white dwarf has been observed at least three times. We used FITSB2 to fit our grid of model spectra to the eight Balmer 
absorption lines ranging from H$\alpha$ to H10 in each exposure. So as to match the instrument resolution, the models were convolved
with a Gaussian with a full width 
half maximum of 0.2 \AA. In addition, points in the observed data lying more than 3$\sigma$ from the model were clipped from subsequent iterations 
of the fitting process. Both the starting values and the new values derived here for $T_{\rm eff}$ and log~$g$ for each white dwarf are given in Table 
\ref{teff}. The errors given in Table \ref{teff} for $T_{\rm eff}$ and log~$g$ are formal fitting errors and are unrealistically small as they neglect systematic
uncertainties e.g. flat fielding errors and model shortcomings. In subsequent discussion 
here we follow \citet{napiwotzki99} and assume an uncertainty of 2.3 per cent in  $T_{\rm eff}$ and 0.07dex in log~$g$. 

\subsection{Determination of line core velocity shifts}

Using our new determinations for the values of $T_{\rm eff}$ and log~$g$ (with the exception of WD0836+201) as input parameters, a model grid which 
was more finely sampled in wavelength space than the first (0.05 \AA) was used in conjunction with FITSB2 to measure the velocity shift of the line
cores of the H$\alpha$ and H$\beta$ lines in the spectra of each white dwarf. Crude estimates for the line core velocity shifts were obtained initially,
taking into account the heliocentric correction. 
The line fitting was then 
re-run using these as starting values for the final fit, to minimise the errors and ensure that each result was robust. We
  made reliable estimates of the uncertainties on the line shift
  measurements based on the the bootstrapping  method of statistical resampling \citep{efron82}.
 We note that a detailed study of 
the wavelengths of prominent water absorption lines in a large number of UVES spectra obtained as part of the SpY project, which uses an identical 
instrumental set-up to this study, indicates that error in the external calibration of the wavelength scale of UVES from spectrum to spectrum is at 
the 0.7 kms$^{-1}$ level. As an independent check of our results we also performed these fits using a different set of synthetic spectra (the LTE model
atmospheres of \citealt{koester01}). Fortunately, no systematic discrepancies were found between the results derived using the two sets of models and 
the differences for any object were well within the measurement uncertainties. The line core velocity measurements for each observation may be found
 in Table \ref{rv}.

\begin{table*}
\caption{\label{rv}ID, line core shift measurements for each data set (RV$_1$,RV$_{2}$, RV$_{3}$, RV$_{4}$, $\chi^{2}$,and log$_{10}$ probability of the white dwarf being a radial velocity variable, for the eight non-magnetic Praesepe white dwarfs investigated. The errors take into account the 0.7 kms$^{-1}$ uncertainty in the UVES wavelength calibration.}
\begin{center}
\begin{tabular}{l  llllcl}
\hline
ID &RV$_{1}$ (kms$^{-1}$)&RV$_{2}$ (kms$^{-1}$)&RV$_{3}$ (kms$^{-1}$)&RV$_{4}$ (kms$^{-1}$)&$\chi^{2}$&log$_{10}$ prob\\ 
\hline
WD0833+194 &  77.13$\pm$2.09 &79.60$\pm$2.15 &83.54$\pm$2.86 &75.69$\pm$2.77&  4.788321&-0.726\\
WD0836+199 &  75.11$\pm$4.88 &85.86$\pm$3.72 &84.66$\pm$2.49 &94.57$\pm$3.49& 11.29552& -1.990\\
WD0837+185 &  70.39$\pm$4.08 &83.44$\pm$1.98 &90.89$\pm$2.51 &-             & 18.70493& -4.062 \\
WD0837+199 &  75.47$\pm$1.94 &76.71$\pm$2.51 &71.25$\pm$2.76 &70.09$\pm$2.12& 5.920174& -0.937\\
WD0837+218 &  80.18$\pm$1.79 &81.95$\pm$2.44 &83.66$\pm$1.88 &-             & 1.955053& -0.425\\
WD0840+190 &  91.32$\pm$7.64 &84.66$\pm$3.10 &82.77$\pm$3.54 &-             & 1.037816&-0.225\\
WD0840+200 &  73.76$\pm$3.27 &75.42$\pm$1.83 &74.82$\pm$2.19 &74.64$\pm$2.24& 0.2153261& -0.011\\
WD0843+184 &  93.40$\pm$3.08 &86.10$\pm$1.99 &87.66$\pm$3.10 &87.80$\pm$2.06& 4.008115& -0.584\\
\hline
\end{tabular}
\end{center}
\end{table*}

In order to describe the statistical
  likelihood that a given target's velocity is constant or otherwise
  between each observation, we calculated a $\chi^{2}$ statistic for each
  star based on the prior assumption that the velocities are stable
  with time i.e. any object with a $\chi^{2}$ below 10 can be considered to have a constant radial velocity. It can be seen from Table \ref{rv} that while the measurements for most objects are consistent with a constant velocity, there is evidence that 
the wavelengths of the line cores in the spectra of WD0837+185 change from exposure to exposure. We note that in the SpY project where velocity measurements
have been obtained for a sample of 1000 white dwarfs, objects with a probability of less than 10$^{-3}$, corresponding to one false "detection" due to 
chance statistical fluctuations, are flagged as potential variables. Given our assessment of the probability of obtaining the tabled radial measurements 
for WD0837+185 by chance ($<$10$^{-4}$), we are led to conclude that this white dwarf is a radial velocity variable.

\section{Results}

\subsection{A possible double degenerate$?$}

\begin{figure}
\begin{center}
\scalebox{0.3}{{\includegraphics[angle=270]{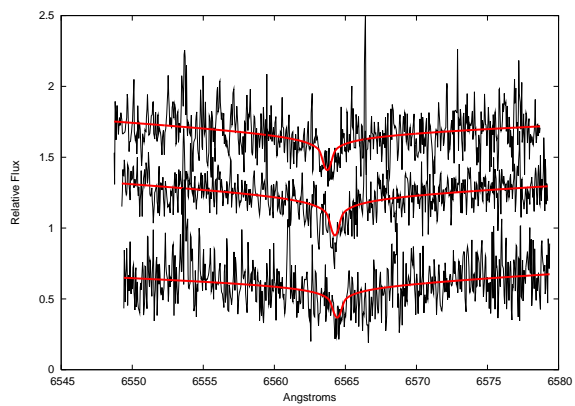}}}
\caption{\label{H_alpha} The H$\alpha$ line profile in the three sets of spectra observed for WD0837+185. The 
red lines are the FITSB2 model fits to the data. The core of the lines can be seen moving. These observations were taken on 2005/11/16 (UT1) and 
2005/12/14 (UT2 + UT3). The observations are plotted in chronological order with the earliest observation at the top of the plot.}
\end{center}
\end{figure}

\begin{figure}
\begin{center}
\scalebox{0.3}{{\includegraphics[angle=270]{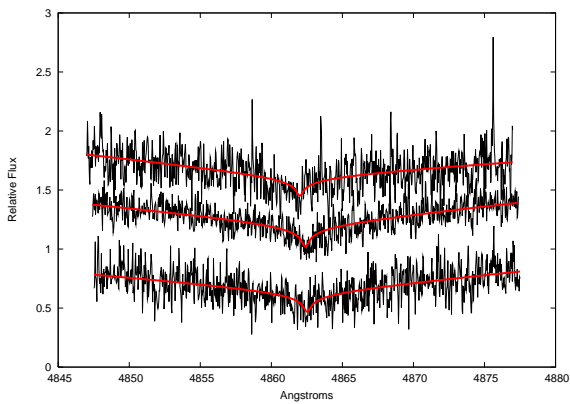}}}
\caption{\label{H_beta}As for Figure \ref{H_alpha}, but for the H$\beta$ line profile from the three sets of spectra observed for WD0837+185.} 
\end{center}
\end{figure}

\begin{figure}
\begin{center}
\scalebox{0.3}{{\includegraphics[angle=270]{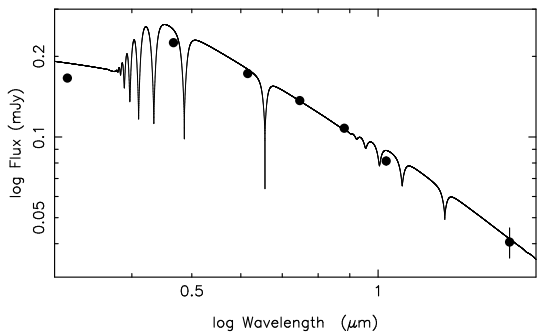}}}
\caption{\label{sdss}Sloan $u$, $g$, $r$ and $i$, and UKIDSS $Z$, $Y$ and $H$ band photometry of WD$0837+185$,
compared to a synthetic DA white dwarf spectrum for $T_{\rm eff} = 15076$K and log~$g = 8.306$. There is no evidence for excess emission due to a cool
companion. At an age of 625Myr these data suggest that such an object, if present, must have a mass $< 0.05$M$_\odot$.} 
\end{center}
\end{figure}

 A cursory glance at Figures \ref{H_alpha} and \ref{H_beta} appears to
substantiate the above conclusion that WD0837+185 is a radial velocity
variable. However, we
find no compelling direct evidence of any cool companion within either the
UVES or the Sloan Digital Sky Survey (SDSS) spectrum of WD0837+185, the
latter extending to
9200\AA. There is also no detection of this object in the 2MASS Point
Source Catalogue \citep{skrutskie06}, although it is listed in the UKIRT
Infrared Sky Survey Large
Area Survey (UKIDSS LAS \citealt{warren07}), with magnitudes
$Z$=18.29$\pm$0.03, $Y$=18.49$\pm$0.04 and $H$=18.50$\pm$0.14 (on a Vega
system). Taken in
conjunction with the SDSS optical photometry ($u$=18.35$\pm$0.03,
$g$=18.02$\pm$0.02, $r$=18.32$\pm$0.01, $i$=18.56$\pm$0.02,
$z$=18.76$\pm$0.04, on approximately an
AB magnitude system), these also provide no evidence for the presence of a
cool companion (see Figure \ref{sdss}). Indeed, based on the DUSTY models
of \citet{chabrier00},
we set a limit of M$_{\rm cool}$$\simless$0.05 M$_{\odot}$, at the
distance and age of this cluster. Given the rarity of brown dwarf + white
dwarf binaries (e.g. \citealt{farihi05} estimate
the fraction of L dwarf companions to white dwarfs to be at $<$0.5 per cent),
this suggests that any companion is more likely to be another white dwarf.

However, there is no evidence in the UVES spectrum which would indicate
the presence of a He-atmosphere DB companion and the observed Balmer line
profiles are relatively well matched
by a synthetic spectrum for a single DA white dwarf. Moreover, at the
spectroscopically determined effective temperature and surface gravity,
the SDSS magnitudes of WD0837+185 are
consistent with those of a single degenerate at the distance of Praesepe.
The models of \citet{holberg06} indicate an isolated DA of this
effective temperature and surface
gravity to have M$_{g}$$\sim$11.6. If we were to assume that WD0837+185
had a degenerate companion with M$_{g}$$\simless$12.5, the absolute
magnitude of this hypothetical binary would
be M$_{g}$$\simless$11.2. Since the observed magnitude of this object is
$g$=18.02$\pm$0.02, this would infer a double degenerate lying well beyond
the back of the cluster (d$\simgreat$220 pc).
So, if this is a binary, the likelihood of Praesepe membership, as
indicated by the proper motion and the mean radial velocity, would appear
to favour a putative white dwarf companion
which is significantly less luminous than WD0837+185.

For a cooling time $\simless$625 Myr (i.e. the age of the cluster) the
available data suggests that any degenerate companion must be massive
(M$\simgreat$1.2 M$_{\odot}$;
M$_{g}$$\simgreat$12.5; \citealt{althaus07, holberg06}),
with a low intrinsic luminosity due to its small radius. If this is the
case, the total mass of this
system would be larger than the Chandrasekhar limit ($\approx$ 1.4
M$_{\odot}$) and this binary has the potential to be a progenitor for a
Type Ia supernova \citep{iben84,
livio00, napiwotzki01, napiwotzki03}. Further study of this system is
warranted to better constrain the masses of the components, and to
determine if the system is close enough that it
will merge within a Hubble time. We note that WD0837+185 sits on the IFMR
defined by the bulk of the other objects (see Figure \ref{IFMR}), so there is
nothing in terms of the mass of this white dwarf
to suggest that this system has experienced close binary evolution.


\subsection{Magnetic white dwarf}
\label{mag}
\begin{figure}
\begin{center}
\scalebox{0.3}{\includegraphics[angle=270]{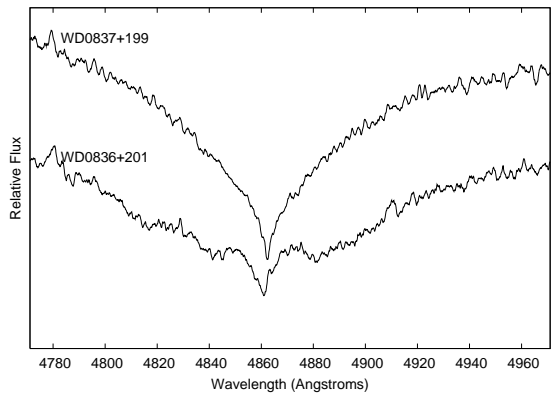}}\\
\caption{\label{wd0836+201} The full median filtered, smoothed, stacked UVES spectra for WD0837+199 and WD0836+201 covering the H$\beta$ line profile. The Zeeman shifted components can be seen for WD0836+201.}
\end{center}
\end{figure}

We note here that the identities of WD0837+199 and WD0836+201 have been confused in some previous studies, e.g. \citet{claver01,
 dobbie04} and \citet{dobbie06}. 
The UVES spectrum of WD0836+201 (LB390, EG59) clearly shows Zeeman shifted components (see Figure \ref{wd0836+201}). This is consistent with \citet{reid96} who 
 notes his fit to the  H$\alpha$ line core in the spectrum of WD0836+201 (LB390) is particularly poor, as 
would be expected if trying to fit Zeeman shifted components with non-magnetic synthetic spectra.
\citet{claver01} also finds a magnetic white dwarf in Praesepe, and in Figure 8 of \citep{claver01}, presents the fit to the spectrum, but labels it as EG61 (i.e. WD0837+199, LB393) for which they determine $T_{\rm eff}$=17,098 K and 
log~$g$=8.32. Moreover, Table 3 of \citet{claver01} erroneously label EG61 as WD0836+201. We believe from our data that the magnetic white dwarf is EG59 and this object has been mislabelled in \citet{claver01}. \citet{dobbie06} and \citet{dobbie04} never observed WD0836+201 or WD0837+199 but refer to EG61 as the magnetic white dwarf as identified by \citet{claver01}. 
The Zeeman shifted Balmer lines in the UVES spectrum of WD0836+201 due to the substantial magnetic field (B$\approx$3 MG), prevents us from
obtaining a meaningful result via fitting the absorption lines with our non-magnetic models.

\begin{table*}
\begin{center}
\caption{ID, name, coordinates, expected line-of-sight velocity
 assuming cluster membership (RV$_{\rm C}$), weighted mean line core shift, gravitational redshift
 based on UVES $T_{\rm eff}$ and log~$g$, line-of-sight velocity
 (RV$_{\rm O}$) and estimated deviation from cluster velocity ((RV$_{\rm O}$-RV$_{\rm C}$)/($\Delta$RV$_{\rm O}$$^{2}$ + 0.8$^{2}$)$^{0.5}$), for the eight non-magnetic white dwarfs
 included in this work.}

\label{wdmass}
\begin{tabular}{llcclcccc}
\hline

ID & Name & RA  & Dec & RV$_{\rm C}$ & H-$\alpha$,H-$\beta$ shift & GR. &
RV$_{\rm O}$$\pm\Delta$RV$_{\rm O}$ & (RV$_{\rm O}$-RV$_{\rm C}$)/($\Delta$RV$_{\rm O}$$^{2}$ + 0.8$^{2}$)$^{0.5}$\\
& & \multicolumn{2}{|c|}{J2000.0} &  \multicolumn{4}{|c|}{kms$^{-1}$} &  \\
\hline

WD0833+194 & --          & 08:36:10.01 & +19:38:19.1 &   35.0 & 78.6$\pm$1.2 & 40.1$\pm$4.4 & 38.5$\pm$4.6 & +0.74\\
WD0836+199 & LB1847,EG60 & 08:39:47.20 & +19:46:12.1 &   34.5 & 86.1$\pm$1.7 & 43.4$\pm$4.7 & 42.7$\pm$5.0 & +1.61\\
WD0837+185 & LB5959      & 08:40:13.30 & +18:43:26.4 &   34.8 & 84.3$\pm$1.5 & 49.2$\pm$5.2 & 35.1$\pm$5.4 & +0.05\\
WD0837+199 & LB393, EG61 & 08:40:28.09 & +19:43:34.8 &   34.5 & 73.5$\pm$1.1 & 41.5$\pm$4.5 & 32.0$\pm$4.6 & -0.53\\
WD0837+218 &             & 08:40:31.47 & +21:40:43.1 &   33.9 & 81.9$\pm$1.1 & 62.8$\pm$6.5 & 19.1$\pm$6.6 & -2.24\\
WD0840+190 &             & 08:42:58.03 & +18:54:35.5 &   34.4 & 84.5$\pm$2.2 & 54.7$\pm$5.7 & 29.8$\pm$6.1 & -0.74\\
WD0840+200 & LB1876      & 08:42:52.32 & +19:51:11.3 &   34.2 & 74.9$\pm$1.1 & 40.1$\pm$4.4 & 34.8$\pm$4.5 & +0.13\\
WD0843+184 & LB8648      & 08:46:01.91 & +18:30:48.5 &   34.2 & 88.0$\pm$1.2 & 51.5$\pm$5.4 & 36.5$\pm$5.5 & +0.41\\
\hline
\end{tabular}
\end{center}

\end{table*}

\subsection{Constraints on the white dwarf mass-radius relation}

The white dwarf mass-radius relation is founded on a Nobel Prize winning theoretical description of the equation of state of an electron 
degenerate gas \citep{chandrasekhar39}. This relation is widely applied in astrophysics and assumed to be robust, yet the empirical evidence in support 
of it is relatively weak (e.g. \citealt{provencal97, provencal02}). This is because, despite it being comparatively straightforward to constrain the radius 
of a white dwarf if its distance is known, it is difficult to reliably or independently determine its mass. For example, for a typical field white 
dwarf with a parallax based distance estimate, the mass can be determined from the radius and the surface gravity, following Equation \ref{eq1}, where $M$ is the 
mass, $g$ the surface gravity, $R$ the radius and $G$ the gravitational constant. However, here the uncertainty in the mass estimate is at least twice the fractional 
error in the radius, resulting in significant scatter amongst empirical data obtained in this way.

\begin{equation} 
\label{eq1}
M = g R^{2} / G
\end{equation}

Fortunately, for white dwarfs which are members of open clusters, both the distance and the radial velocity are constrained and thus it is possible to determine 
the gravitational redshift. The masses of these objects can then be estimated via Equation \ref{eq2}.
\begin{equation}
 \label{eq2}
M = v  R / 0.635, 
\end{equation}
where $v$ is the gravitational redshift in kms$^{-1}$ and $M$ 
and $R$ are mass and radius in solar units respectively.
 The uncertainty in the mass estimate is reduced since the fractional error in the radius 
determination has half the impact as in Equation \ref{eq1}.

In light of this, we have assumed that the candidate Praesepe white dwarfs which are consistent with the general trend of the IFMR, as defined by data points 
from other systems, are robust cluster members. In this case, the distance to each degenerate has been taken
to be that to the centre of the Praesepe, but with an uncertainty associated with the intra-cluster spatial distribution of the white dwarfs. Here, we 
adopt D=184.5$\pm$6 pc as the distance to the cluster centre, which is the weighted mean of the Hipparcos based measurement, (m-M)$_{0}$=6.24$\pm$0.12 
\citep{mermilliod97}, the ground based parallax measurement of \citep{gatewood94}, (m-M)$_{0}$=6.42$\pm$0.33 and a recent photometric determination which 
takes into account a new spectroscopic determination of the cluster metallicity, (m-M)$_{0}$=6.33$\pm$0.04 (\citealt{an07}). All white dwarf candidate members of Praesepe
identified in the survey of \citet{dobbie04}, which extended beyond 2.5$^{\circ}$ of the cluster, are found to lie within 2$^{\circ}$ of the cluster centre, corresponding 
to 6.5 pc at D=184.5 pc. If the distribution of the white dwarfs is spherically symmetric about the cluster centre then all are likely to lie within the distance range 
D=184.5$\pm$8.5 pc. 

\begin{figure}
\begin{center}
\vspace{0.5cm}
\scalebox{0.95}{\includegraphics[angle=0]{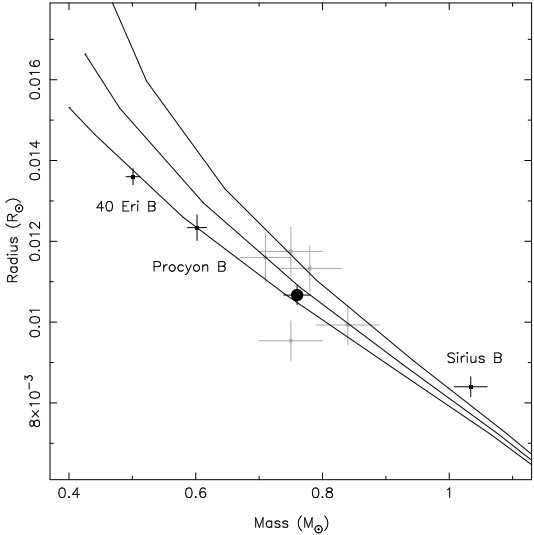}}
\caption{\label{MRR}The mass-radius relation for white dwarfs. The locations of the five individual Praesepe white dwarfs for which we 
were able to estimate radius and mass are shown (light-grey points) together with their weighted mean (large black circle). Three degenerate 
components of visual binary systems for which the most robust estimates of mass and radius are currently available (black squares) and 
theoretical mass-radius relations for white dwarfs with CO cores with thick H layers at effective temperatures of $T_{\rm eff}$=5000 K (bottom), 
15000 K (middle) and 30000 K (top) are also shown (solid black lines).
}
\end{center}
\end{figure}

The radius of each degenerate has been determined by scaling the distance (D) by the square root of the estimated flux ratio, (f/F)$^{0.5}$, 
where f is the observed flux at the Earth's surface and F is the flux at the surface of the white dwarf. The surface flux in the Sloan Digital Sky Survey 
(SDSS) filters ($r$, $i$ and $z$; e.g. \citealt{fukugita96}) for each star has been obtained from the pure-H model atmospheres of \citep{holberg06}, where 
 cubic splines have been used to interpolate between points in the grid. The uncertainty in the surface flux due to the error 
in the surface gravity determination for each object was calculated to be 0.3-0.4 per cent, assuming a generous uncertainty of $\pm$0.1 in the log~$g$ measurement, and so was assumed negligible. However, in the temperature range of interest, the error in the 
effective temperature determination for each object was found to introduce an uncertainty of $\sim$3.5 per cent in F ($r$, $i$ and $z$). We note that this 
uncertainty would have been larger if the $u$ or $g$ bands were used. The observed SDSS fluxes (Table \ref{mass/rad}) are believed to have absolute uncertainties at the 2 per cent 
level \citep{adelman08}. Thus the estimated total combined uncertainty in our white dwarf radius determinations is 
$\sim$5 per cent, where contributions from errors in the white dwarf distances, the predicted fluxes and the observed fluxes are 4.6 per cent, 1.75 per cent and 1 per cent 
respectively and have been added quadratically.

Subsequently, the mass for each star has been re-calculated using Equation \ref{eq2}, this time using a gravitational redshift determined by subtracting the predicted
line-of-sight velocity (RV$_{\rm C}$ in Table \ref{wdmass}), assuming each white dwarf shares the cluster motion, from the weighted mean measured shift of the non-LTE cores of the 
H$\alpha$ and H$\beta$ lines (Table \ref{mass/rad}). As open cluster escape velocities are typically low, approximately twice the  velocity dispersion (e.g. \citealt{fellhauer03}), any object that has a space velocity substantially different from the cluster mean will be rapidly lost from the system. We note in this 
context that as the white dwarf members of Praesepe have cooling ages of $\tau$$_{\rm cool}$$\sim$200-300 Myr, it is unlikely that putative point asymmetries
in mass loss during the later stages of AGB evolution have displaced their space velocities greatly from the cluster mean.
Moreover, the velocity dispersion of $\sim$1 M$_{\odot}$ cluster members as determined by \citet*{mermilliod99} 
is only 0.8 kms$^{-1}$. Propagating these uncertainties, we estimate that the mass determination for each white dwarf using the gravitational redshift should be robust to $\sim$6  per cent. 

The magnetic white dwarf, WD0836+201, the possible double-degenerate system, WD0837+185, WD0836+199 which has no clean SDSS photometry and WD0837+218 
which sits well above the bulk of objects in the IFMR have been omitted from these calculations. Our estimates of the masses and radii of the remaining 
five white dwarfs \begin{table*}
\begin{center}
\caption{\label{mass/rad}The SDSS magnitudes, the model independent masses and radii for the white dwarf members of Praesepe with UVES 
spectroscopy. The magnetic white dwarf, WD0836+201, the possible double-degenerate system, WD0837+185, WD0836+199 which has no clean SDSS photometry and WD0837+218, which sits above the bulk of white dwarfs in 
the IFMR, have been omitted from these calculations.}
\begin{tabular}{l cccc c}
\hline
Name & $r$ & $i$ & $z$ &$M$& $R$\\
&&&&M$_{\odot}$&$\times$10$^{-2}$ R$_{\odot}$\\
\hline
WD0833+194 &18.09$\pm$0.02 &18.34$\pm$0.02 &18.65$\pm$0.04 &0.78$\pm$0.05 &  1.133$\pm$0.057\\
WD0837+199 &17.81$\pm$0.01 &18.08$\pm$0.02 &18.34$\pm$0.04 &0.71$\pm$0.04 &  1.160$\pm$0.059\\
WD0840+190 &18.44$\pm$0.01 &18.70$\pm$0.02 &18.98$\pm$0.04 &0.75$\pm$0.05 &  0.954$\pm$0.049\\
WD0840+200 &18.01$\pm$0.01 &18.21$\pm$0.02 &18.57$\pm$0.05 &0.75$\pm$0.05 &  1.175$\pm$0.060\\
WD0843+184 &18.36$\pm$0.01 &18.61$\pm$0.02 &18.85$\pm$0.04 &0.84$\pm$0.05 &  0.993$\pm$0.050\\
\hline
\end{tabular}
\end{center}
\end{table*}are given in Table \ref{mass/rad}.

These masses and radii are plotted in Figure \ref{MRR} (light grey points with error bars) together with the most robust points currently available 
in the literature (small points with error bars). The masses of Procyon B ($T_{\rm eff}$$\approx$7700 K), 40 Eri B ($T_{\rm eff}$$\approx$16700 K) and Sirius B 
($T_{\rm eff}$$\approx$25000 K) have been determined precisely using dynamical methods since these white dwarfs are members of visual binaries. Moreover, the 
distances to these three systems are well constrained by Hipparcos parallax measurements \citep{esa97}. Theoretical mass-radius relations for  $T_{\rm eff}$=5000, 15000 
and 30000 K, based on the CO core, thick H-layer evolutionary models of \citet*{fontaine01} are also shown (black lines). Four out of the five 
Praesepe stars are located within 1$\sigma$ of the $T_{\rm eff}$=15000 K track. WD0840+190 lies somewhat below this ($\sim$2$\sigma$). It is possible that the 
radius of this object has been underestimated perhaps because it resides on the very back edge of the cluster. However, there is good agreement between the 
spectroscopic and flux based radius estimates for all five white dwarfs. Alternatively, the radial velocity of this object may be overestimated by $\sim$4-5 kms$^{-1}$, resulting in a marginally low gravitational redshift determination. WD0840+190 may have very recently received a velocity kick through an interaction 
with a binary system near the centre of the cluster and be in the process of leaving Praesepe. A remaining possibility is that the low-level ripple in some of 
our UVES spectra could have adversely affected our determination of the shift of the line cores.  Given that the slope of the IFMR is not expected to change significantly
within the comparatively narrow mass window in which our objects reside,
the impact of a number of uncertainties e.g. the velocity
 dispersion, the intra-cluster distribution of the white dwarfs and the
 random errors in our line core shift measurements, can be reduced if
 we take the weighted mean of the masses and the radii of the five stars.In this case, we determine the $M_{\rm mean}=0.760\pm0.021 M_{\odot}$ and $R_{\rm 
mean}$$=0.01067\pm0.00024$~$R_{\odot}$ (large filled circle in Figure \ref{MRR}). 

Given that the mean effective temperature of these five objects is $T_{\rm eff}$$\approx$15400 K 
we find the location of this point in mass-radius space to be entirely consistent with the theoretical relation. Indeed, the proximity of all four robust data points
to the model tracks, strongly supports the theoretical white dwarf mass-radius relation and confirms the veracity of the prediction that white dwarfs with larger 
masses have smaller radii.

\subsection{Radial velocities and the membership status of WD0837+218}
\label{gr}

As has been discussed in the introduction, the white dwarf members of open clusters are extremely useful for constraining the form of the IFMR. 
However to assess the level of intrinsic scatter in this relation it is crucial that field white dwarf interlopers are eliminated from samples. 
An excellent way to adjudge cluster membership is via the measurement of radial velocities. Open star clusters typically have small velocity 
dispersions. For example, \citet*{mermilliod99} determine a velocity dispersion of only 0.8 kms$^{-1}$ for a sample of single F5-K0 members
of Praesepe. Moreover,  as discussed previously, it is unlikely that these white dwarfs will have velocities which differ substantially from 
the mean cluster velocity due to evolutionary effects (e.g. asymmetric mass loss) since they have remained within the projected tidal radius of 
Praesepe despite having formed 200-300 Myr ago.
While the gravitational redshift component to the velocity shift of the absorption line cores in the spectra of white dwarfs is appreciable, 
given the robustness of evolutionary models as demonstrated above, if the effective temperature and surface gravity of a star is known, then the 
magnitude of this effect can be calculated straightforwardly using Equation \ref{eq2}.

Hence the radial velocity for each candidate Praesepe degenerate has been obtained by taking the difference between the measured shift of the 
line cores and the gravitational redshift as determined from evolutionary models and the measured gravity and effective temperature. It can be seen from 
Table \ref{wdmass} that, as expected, the radial velocities of the bulk of the white dwarfs, 
including the mean velocity of our candidate double-degenerate system, lie within 1$\sigma$ of that expected if they are members of Praesepe. 
These latter values have been estimated assuming a total cluster space velocity of V$_{\rm TOT}$=47.5 kms$^{-1}$, 
a convergent point of RA=06h14m00s, Dec=-04d36m00s, J2000.0 (e.g. \citealt{mermilliod90, mermilliod99}) and the method detailed in \citet{reid96}. 
We find that the measured radial velocity of only one 
star,  WD0837+218, is notably different (14.8$\pm$6.6 kms$^{-1}$ or 2.24$\sigma$ below), from the velocity predicted on the assumption of cluster
membership. There is no evidence in the available data that this object is a radial velocity variable. As this white dwarf has the largest projected 
separation from the cluster centre of those identified by \citet{dobbie04}, and appears to be an outlier in the IFMR of \citet{dobbie06} we 
are led to conclude that it is more likely a field star interloper than a member of Praesepe. Assuming this to be correct, it should not be used for 
constraining the form of the IFMR.

\subsection{Minor revision to the Initial Mass-Final Mass Relation}

\label{ifmr}
These new UVES data for eight white dwarf members of Praesepe are only of moderate S/N per pixel but are homogeneous in nature and 
of much higher resolution than previously published spectra. The line shift measurements suggest that previous studies of 
these stars may have systematically underestimated by a small amount ($\sim$0.1) their surface gravities. For example, for WD0843+184 
we derive a radial velocity of 45.2 kms$^{-1}$ based on the surface gravity (and effective temperature) of \citet{dobbie06}. 
Additionally, we estimate a radial velocity of 42.9 kms$^{-1}$ for WD0837+199 adopting the surface gravity (and effective temperature) 
erroneously assigned to WD0836+201 in \citet{claver01}. On this basis we argue that our measurements of effective 
temperature and surface gravity are, except in the case of the magnetic white dwarf WD0836+201, to be preferred to those 
obtained previously.
  
As in our earlier work (e.g. \citealt{dobbie06}), we have utilised a grid
of evolutionary models based on a mixed CO core
composition and a thick H surface layer (e.g. \citealt{fontaine01}) to
estimate the mass and cooling time
of each white dwarf from our measurements of effective temperature and
surface gravity (see Table \ref{teff}). Cubic splines have
been used to interpolate between the points within this grid. The lifetime
of the progenitor star of each white dwarf has
then been calculated by subtracting the cooling time from the age of the
cluster (625$\pm$50 Myr). Subsequently, to constrain
the mass of the progenitor star, as in our earlier work we have used the stellar evolution
models of \citet{girardi00}, here for solar metallicity (Z$_{\odot}$=0.019, which is close to the
 value favoured by \citealt*{anders89}). Again cubic splines were used to interpolate between 
the points in the grid. We find that these model grids indicate that the difference between the
lifetimes of 3-4 M$_{\odot}$ stars
with Z=0.019 and Z=0.030 compositions is only $\approx$1 per cent. As the
metallicity of Praesepe is midway between these values, the use of calculations
for solar composition to estimate initial masses here appears adequate.
Our estimates of the masses, cooling times and progenitor
masses for each of the white dwarfs are shown in Table \ref{teff}.

\begin{figure*}
\begin{center}
\scalebox{0.5}{\includegraphics[angle=270]{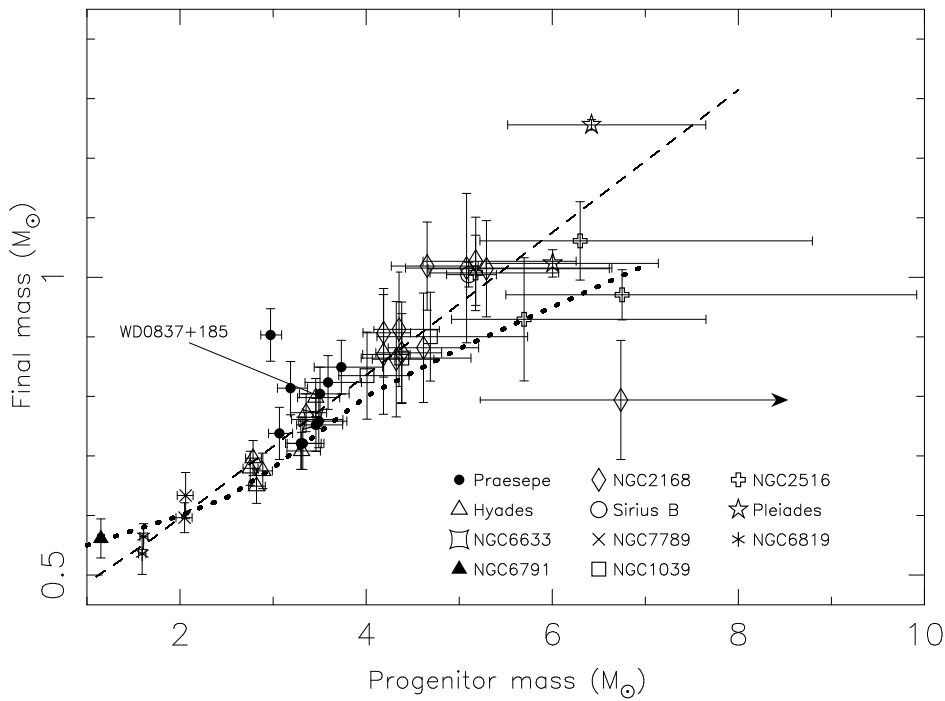}}
\caption{\label{IFMR}A plot showing the revised locations of the Praesepe white dwarfs in initial mass-final mass 
space (filled circles). The locations of the degenerate members of the NGC6633 (square star), NGC7789 (cross), NGC6819 (asterisk),
Hyades (open triangles), Sirius binary system (open circle), the Pleiades (open stars), NGC2168 (open diamonds), NGC1039 (open squares), NGC6791 (filled triangles)
 and NGC2516 
(open '+' signs) are also shown. The relation of \citet{weidemann00} (dotted line) and a revised linear fit to 41 white dwarfs 
(dashed line) are overplotted. The radial velocity variable, WD0837+185 is labelled.}
\end{center}
\end{figure*}

The revised locations of the Praesepe white dwarfs are plotted in Figure \ref{IFMR}, where the cooling times, masses and initial masses 
for WD0836+197 (LB5893), WD0840+205 and WD0836+201 are based on effective temperature and surface gravity estimates from the 
literature \citep{claver01, dobbie06}. WD0837+218 has been excluded since based on the arguments presented above it appears more 
likely to be a field star. We also include in this plot the hydrogen rich white dwarf members of a number of other systems where, for a given 
initial mass range, parameters such as cluster age are relatively well constrained -  namely  NGC6819 and NGC7789 \citep{kalirai08}, NGC6791 \citep{kalirai07}, 
NGC6633 \citep{williams07}, the Hyades \citep{bergeron95b}, NGC2168 (M35; \citealt*{williams04, williams08}), NGC2516 
\citep*{koester96} and the Pleiades \citep{bergeron95, dobbie06, dobbie06b} and the Sirius binary \citep{liebert05b}. For the degenerate members of the 
Hyades, NGC2168, NGC2516, the Pleiades and the Sirius binary we have made the same assumptions 
regarding white dwarf effective temperatures and surface gravities and cluster ages and metallicities as we did in \citet{dobbie06, dobbie06b}. For NGC6819,
 NGC7789 and NGC6633 we have assumed solar metallicity and have adopted the cluster ages and white dwarf effective 
temperatures and surface gravities given in the referenced literature. We
have excluded the three lower mass white dwarf candidate members of
NGC1039 \citep{rubin08} and NGC7063 LAWDS 1  \citep{williams07} since proper motion measurements indicate
these are probably field stars \citep{dobbie09}.  For the
super-solar metallicity NGC6791 we have used the [Fe/H]=+0.18 models of
\citet{girardi00} since this is the most metal rich composition available in their grid.  We have included only the white 
dwarfs with a mass indicative of a CO core. The lower mass helium
white dwarfs in this cluster could be the products of close
binary evolution (e.g. \citealt{bedin08, vanloon08}) or the products of a
metallicity enhanced mass loss mechanism \citep{kalirai07}.
Note that as in \citet{dobbie06, dobbie06b}, the NGC2168 white 
dwarf candidate member LAWDS 22 was not included here as \citet{williams04}  were unable to obtain a satisfactory model 
fit to the observed Balmer lines.

As has been shown by a number of authors (e.g. \citealt{ferrario05, dobbie06, williams07, kalirai08}), the 
IFMR as defined by the vast majority of current semi-empirical data (Figure \ref{IFMR}) can be reasonably approximated by a simple linear 
function (dashed line). Interestingly, WD0837+185, the radial velocity variable, does not appear to be an outlier. If the system experienced
a common envelope phase then it may well have occurred during the later stages of the AGB. In this case the evolution would approximate that
of a single star. Indeed, there is only one star, WD0836+197, which deviates substantially from the general trend. Intriguingly, there 
appears to be nothing else untoward about this star. For example, \citet{dobbie06} confirmed it to be a proper motion member, while the 
spectroscopy of \citet{claver01} reveals no evidence of Zeeman shifts in the Balmer line cores. The line-of-sight velocity 
as derived from the line core velocity shift measurement of \citet{reid96} and the spectroscopic effective temperature and surface gravity 
estimates of \citet{claver01} is extremely close to the value expected on the basis of cluster membership. This argues against, but does not 
rule out, it being a radial velocity variable. Furthermore, \citet{karl05} found no evidence for rapid rotation, the shape of the H$\alpha$ line core 
setting a limit of $v$ sin $i$ $<$22 kms$^{-1}$. 

The magnetic white dwarf, WD0836+201, sits marginally above the bulk of white dwarfs in Figure \ref{IFMR}. However, $T_{\rm eff}$ and log~$g$ 
from line fitting are likely to be less robust here than for non-magnetic degenerates. Indeed, the SDSS photometry ($g-i$, $u-r$) is more consistent with a 
slightly lower effective temperature ($T_{\rm eff}$$\sim$15000 K). We note that despite a gravitational redshift based mass estimate \citep*{heber97} appearing to be consistent with the spectroscopic mass determination of \citet{claver01}, this result is void due 
to the confusion between WD0836+201 and WD0837+199. In any case, it is not obvious that a strongly magnetic white dwarf should be expected to 
follow an IFMR delineated by non-magnetic stars. Magnetism may hinder mass loss during post-main-sequence evolution \citep*{wickramasinghe00}. Alternatively, all strongly magnetic white dwarfs (B$\simgreat$1 MG) might be produced through close binary interaction. \citet{tout08}
propose that isolated high field magnetic white dwarfs form via the merging of two stellar cores within a common envelope environment. This 
unusual evolutionary history would presumably result in a higher than expected white dwarf mass for the estimated progenitor mass. 

Thus the location of WD0836+197 in initial mass-final mass space may represent the best evidence for strong differential mass loss. However 
the lack of other comparatively deviant points in Figure \ref{IFMR} and the sharply peaked form of the white dwarf mass distribution leads us to 
still prefer an evolutionary scenario in which it has formed from a blue straggler star. Praesepe is known to contain a number of blue straggler 
stars (e.g. 40 Cancri and Epsilon Cancri; \citealt*{ahumada07}).

 A revised linear least squares fit to the 41 white dwarfs in this plot, 
which excludes the magnetic white dwarf WD0836+201 and the outlier WD0836+197, between progenitor masses of 1.15 M$_{\odot}$ and  7 M$_{\odot}$ gives parameters m=0.1197$\pm$0.0074 and c=0.3569$\pm$0.0220. These 
parameters are not greatly different from our original estimates of m=0.133$\pm$0.015 and c=0.289$\pm$0.051 \citep{dobbie06} or those 
from the recent work of other groups e.g.  \citet{kalirai08} determine m=0.109$\pm$0.007 and c=0.394$\pm$0.025. 

\section{Summary}

We have presented an analysis of high resolution VLT and UVES spectroscopy of nine white dwarf candidate members of Praesepe. For the eight non-magnetic stars we have 
measured the velocity shift of the H$\alpha$ and H$\beta$ line cores and obtained revised estimates of their effective temperatures and surface gravities.  We 
have shown that contrary to the conclusions of a number of previous studies, WD0836+201 is the magnetic white dwarf. We have also provided evidence that 
WD0837+185 is a radial velocity variable and possibly a double-degenerate system. Additionally, we have used the five most appropriate Praesepe stars to obtain 
a new robust constraint on the white dwarf mass-radius relation. This new data point and the three most robust existing data points have been shown to lie in close 
proximity to the theoretical tracks. This result confirms the veracity of modern theoretical white dwarfs mass-radius relations. In light of this we have 
determined the radial velocities of all eight non-magnetic degenerates. We have found that the line-of-sight velocity of WD0837+218, 
when examined in conjunction with its projected position with respect to the cluster centre and its location in initial mass-final mass space, argues that it is more 
likely to be a field star than a cluster member. It should thus be discounted from investigations concerned with the form of the initial mass-final mass relation. 
We have employed our new parameters for the Praesepe white dwarf members, in conjunction with comparatively robust data from other open clusters to re-assess the form 
of the initial mass-final mass relation. We find that there is only one substantial outlier, WD0836+197, from a near monotonic IFMR, which can still be adequately 
represented by simple linear function. The location of this star in initial mass-final mass space may be evidence of strong differential mass loss but given the lack 
of other outlying stars and the sharply peaked nature of the white dwarf mass distribution, we still favour an evolutionary scenario in which it has formed from a 
blue straggler star. 

%
\section*{Acknowledgements}
SLC acknowledges the financial support of STFC. RN and MBU are supported by STFC Advanced Fellowships.
This research has made use of NASA's Astrophysics Data System. Funding for the creation and distribution of the SDSS Archive has 
been provided by the Alfred P. Sloan Foundation, the Participating Institutions, the National Aeronautics and Space Administration, 
the National Science Foundation, the U.S. Department of Energy, the Japanese Monbukagakusho, and the Max Planck Society. The SDSS Web 
site is http://www.sdss.org/.he SDSS is managed by the Astrophysical Research Consortium for the Participating Institutions. The Participating Institutions are the American Museum of Natural History, Astrophysical Institute Potsdam, University of Basel, University of Cambridge, Case Western Reserve University, University of Chicago, Drexel University, Fermilab, the Institute for Advanced Study, the Japan Participation Group, Johns Hopkins University, the Joint Institute for Nuclear Astrophysics, the Kavli Institute for Particle Astrophysics and Cosmology, the Korean Scientist Group, the Chinese Academy of Sciences (LAMOST), Los Alamos National Laboratory, the Max-Planck-Institute for Astronomy (MPIA), the Max-Planck-Institute for Astrophysics (MPA), New Mexico State University, Ohio State University, University of Pittsburgh, University of Portsmouth, Princeton University, the United States Naval Observatory, and the University of Washington. 
 This work is based in part on data obtained as part of the UKIRT Infrared Deep Sky Survey.
%
%

\bibliographystyle{mn2e}
%

\bibliography{mnemonic,refs}
\label{lastpage}

\end{document}